\documentstyle[prl,multicol,epsf,epsfig,aps]{revtex}
\newcommand{\BEQ}{\begin{equation}}
\newcommand{\EEQ}{\end{equation}}
\newcommand{\BEA}{\begin{eqnarray}}
\newcommand{\EEA}{\end{eqnarray}}

\renewcommand{\d}{{\rm d}}

\newcommand{\SN}{S_{{\rm vN}}}

\newcommand{\p}{{\partial}}
\newcommand{\Q}{{\cal Q}}

\newcommand{\half}{\frac{1}{2}}
\renewcommand{\thesection}{\arabic{section}}

\def\dbarrm {{\mathchar'26\mkern-11mu{\rm d}}}                       %
\begin{document} 
\draft
\title
{Breakdown of the Landauer bound for information erasure 
in the quantum regime} 
\date{\today}
\author{A.E. Allahverdyan$^{1,2,4)}$ and Th.M. Nieuwenhuizen$^{3)}$}
\address{$^{1)}$ S.Ph.T., CEA Saclay, 91191 Gif-sur-Yvette cedex, France;\\
$^{2)}$ Institute for Theoretical Physics,
$^{3)}$ Department of Physics and Astronomy,\\ 
University of Amsterdam,
Valckenierstraat 65, 1018 XE Amsterdam, The Netherlands; \\ 
$^{4)}$Yerevan Physics Institute,
Alikhanian Brothers St. 2, Yerevan 375036, Armenia. }
\maketitle

\begin{abstract}
A known aspect of the Clausius inequality is that 
an equilibrium system subjected to a squeezing $\d S$ of its
entropy must release at least an amount $|\dbarrm Q|=T|\d S|$ of heat.
This serves as a basis for the Landauer principle, which puts 
a lower bound $T\ln 2$ for the heat generated by erasure of 
one bit of information. Here we show that in the world of quantum
entanglement this law is broken. 

A quantum Brownian particle 
interacting with its thermal bath can either generate less heat
or even {\it adsorb} heat during an analogous squeezing process,
due to entanglement with the bath. The effect exists even for weak
but fixed coupling with the bath, provided that temperature 
is low enough. This invalidates the Landauer bound in the quantum 
regime, and suggests that quantum carriers of information 
can be much more efficient than assumed so far.

\end{abstract}


\renewcommand{\thesection}{\arabic{section}}
\section{ Introduction}
\setcounter{equation}{0}\setcounter{figure}{0} 
\renewcommand{\thesection}{\arabic{section}.}

The laws of thermodynamics are at the basis of our
understanding of nature, so it is rather natural that 
they have many application much beyond their original 
scopes, e.g. in computing and information processing
\cite{rex,Neuman,leo,landauer,landauer1,ben,zurek,japan}. 
The first connection between information storage and thermodynamics 
was made by von Neumann in the 1950's~\cite{Neuman}. His speculation 
that each logical operation costs at least an amount of energy 
$T\ln 2$ proved too pessimistic. Landauer pointed out that 
reversible ``one-to-one'' operations can be performed, in principle,
without dissipation; only irreversible operations ``many-to-one'' 
operations, like erasure, require dissipation of energy, 
at an amount at least equal to the von Neumann estimate 
$T\ln 2$ per erased bit~\cite{landauer,landauer1}. 
This conclusion is a direct
consequence of the Clausius inequality, which connects the 
change of heat in a given process with the change of entropy.
It is perfectly consistent with intuition, since as 
everybody had a chance to observe, equilibrium substances 
typically release heat under isothermal compression of their 
entropy (or volume, which is the same for a good majority of classical
cases). 
Rather recently the base of the effect was finally put on the
Clausius inequality \cite{japan}, and it was shown that the previous
not very strict considerations are particular cases of its 
application to information processing systems.

The principal importance of erasure among other information-processing
operations originates from the fact that it is connected with changes of 
entropy, and thus cannot be realized in a closed system. One needs to couple
the information-carrying system with its environment. Therefore the process 
is accompanied with changes in heat whose magnitude
has to be determined by thermodynamics. 
It was shown rigorously that all computation can be performed using reversible
logical operations only~\cite{ben}.

Here we will consider thermodynamical aspects of erasure in the 
context of low temperatures, so low that the quantum effects start 
to play an 
important role. We choose the simplest working example: One-dimensional
Brownian particle in contact with a thermal bath at temperature $T$,
and in the presence of an external confining potential.
The main new aspect arising at low temperatures is 
entanglement of the Brownian particle with the bath.
Therefore, even when the total system is in a pure state, the
subsystem (Brownian particle) is in a mixed state.
Thus its stationary state cannot be given by equilibrium 
quantum thermodynamics, which would predict for $T\to 0$ that the 
subsystem goes to its groundstate, the pure vacuum state.
The latter can only be reached for not too low temperatures, given 
a fixed but weak coupling with the bath. In general
a new situation arises, for which a generalized
thermodynamical interpretation can be given
\cite{AN,AN1} (for analogous situations in glasses
and related systems see \cite{N1,AN0}). 
In particular, the classical Clausius inequality is invalid
and the classical intuition fails. We stress that
this situation is not at all exceptional, since it appears even
for a small but generic coupling if temperature is low enough.

Our main result will show that when entropy of 
the particle is decreased by external agents, 
namely a part of information carried by it is erased,
the particle can 
{\rm adsorb heat} in a clear contrast with the classical intuition.
Later we shall apply this result to show that there is not anything
similar to Landauer bound at low temperatures. Thus in this respect 
quantum carriers of information can be more efficient than
their classical analogs.  

Since we are in a new situation where relations of the 
standard thermodynamics are possibly broken, we prefer 
to work with simple exactly-solvable models, where all
general relations can be illustrated or disproved explicitly.
In analogous situation with the classical theory Szilard 
\cite{leo} used a model with one classical 
Brownian particle interacting with its thermal bath.

This paper is organized as follows. In section 2 we review
the connection between thermodynamics and information erasure. 
Section 3 is devoted to heat and entropy changes of a
quantum Brownian particle in contact with its
thermal bath. This model can be considered as an extension to the 
quantum regime of the seminal model by Szilard \cite{leo}.
In section 4 our main results on violation of the Landauer principle
are presented. In section 5 we analyze the most popular derivation
of this principle \cite{landauer,landauer1,ben} in order to 
show where its arguments appear to be inapplicable.
Our conclusions are presented in the last section.

\renewcommand{\thesection}{\arabic{section}}
\section{Erasure of information and Gibbsian thermodynamics}
\setcounter{equation}{0}\setcounter{figure}{0} 
\renewcommand{\thesection}{\arabic{section}.}

\subsection{Source of information}
Let us start by briefly recalling what is meant by erasure of information.
Since information is carried by physical systems, messages are
coded by their states, namely every state (or possibly group
of states) corresponds to a ``letter''. The simplest example
is a two-state system, which carries on one bit of information.
The basic model of {\it source of information} in Shannonean, 
probabilistic information theory \cite{inf,inf1,jaynes} assumes that
the carrier of information can be in different states with certain
(so called a priori) probabilities. In other words, the messages of this
source appear randomly and the measure of their expectation is given 
by the corresponding probabilities. For example, in the classical case
the carrier of information may occupy a cell in its phase space with volume
$\d x\,\d p/(2\pi\hbar )$ and a priori probability $P(x,p)$.
Then different cells will correspond to different messages.
In the quantum case the completely analogous situation is described by 
a density matrix $\rho $,
\BEA
&& \rho =\sum_n p_n |n\rangle \langle n|,\\
&&\langle n|m\rangle = \delta _{nm},
\label{tiger}
\EEA
which means that the carrier occupies a state $|n\rangle$ with the
a priori probability $p_n$. Moreover, different quantum states are exclusive
as indicated by Eq.~(\ref{tiger}). As in the classical case, the appearance
of the carrier in different states will bring different messages.

The fundamental theorem by Shannon \cite{inf,inf1,jaynes} 
states that the information carried by an information source 
is given by its entropy. Namely it is equal to 
\begin{eqnarray}
\label{w22}
&&S=-\int\frac{ \d x~\d p}{2\pi \hbar}
P(p,x)\ln P(p,x)
\end{eqnarray}
in the classical case, and to
\BEA
\label{q2}
\SN (\rho )=-\sum_{n}p_n\ln p_n = -{\rm tr}(\rho \ln \rho )
\EEA
in the quantum situation.
Here $\SN (\rho )$ is the von Neumann entropy of the density matrix $\rho$.
The physical meaning of this result 
can be understood as follows. A source which has less entropy
occupies less states with higher probability. It can be said 
to be more known, and therefore the appearance of its results
will bring less information. In contrast, a source with higher 
entropy occupies more states with lower probability. Its messages
are less expectable, and therefore bring more information.
The rigorous realization of this intuitive arguments appeared to 
be the most straightforward and fruitful proof of the Shannon 
theorem \cite{jaynes,balian}.
Notice that the entropies (\ref{w22}, \ref{q2}) appear here on 
the information theoretical footing and not as purely 
thermodynamical quantities \cite{jaynes}.

Needless to mention that the above notion of information source 
does not exhaust the full meaning of this concept. Here it 
appears as a model of the probabilistic information theory.
Advantages and shortcomings of this approach were nicely reviewed
by Kolmogorov \cite{kolmo}.

\subsection{Erasure}

Erasure is an operation which is done by an external agent in order
to reduce the entropy of the carrier of information. This means 
that in its final state the carrier brings less information, i.e. 
some amount of it has been erased. 
In particular, the complete erasure corresponds to the minimization 
of entropy. Notice that the erasure is defined as a ``blind'' operation, 
which is done independently on the actual state of the information carrier.
This is how information processing systems operate, they do not recognize
the actual state of a bit before to erase it. Following to standard
assumptions \cite{klim,balian} we will model external operations by a
time-dependent Hamiltonian $H(t)$ of the carrier, namely some of its 
parameters will be varied with time according to given trajectories.
If the information carrying system is closed, then its dynamics is
described by the Liouville equation for $P(x,p)$ in the classical case
or by the von Neumann equation 
\BEA
\label{kro}
\frac{\d}{\d t}\rho =\frac{i}{\hbar}[\,\rho (t)H(t)-H(t)\rho (t)\,]
\EEA
in the quantum situation. As can be shown directly, the entropies
(\ref{w22}, \ref{q2}) remain constant with time. In order to 
change them, one has to consider an information carrier, 
which is an open system. In that
case a part of its energy will be controlled (i.e. 
transferred to or received) by its environment as {\it heat}. 
Indeed, if 
\BEA
\label{w2}
U={\rm tr}[\,H(t)\rho (t)\,]
\EEA 
is the average energy of the carrier, then its change during time 
$\d t$ reads:
\begin{equation}
\label{dE}
\d U=\dbarrm {\cal Q}+\dbarrm {\cal W}
    = {\rm tr}[\,H\d \rho \,] +{\rm tr}[\,\rho\, \d H \,]
\end{equation}
This is the energetic budget of the system.
The last term is the averaged mechanical work
$\dbarrm {\cal W}$ produced by an external agent \cite{klim,balian}.
The first term in r.h.s. of Eq.~(\ref{dE}) arises
due to the statistical redistribution in 
phase space. We shall identify it with the change of heat 
$\dbarrm {\cal Q}$ \cite{klim,balian}, 
so Eq. (\ref{dE}) is just the first law. As can be shown through
Eq.~(\ref{kro}), the heat is explicitly zero for a closed system.
All these formulas are valid in the classical case as well. Here
$\rho $ should be substituted by $P(x,p)$, and the trace will
be changed by the integration over the corresponding phase space.

\subsection{Brownian particle as an information carrier}

In order to specify the situation, let as consider a Brownian particle
as an information carrying system. A similar simple model was employed 
by Szilard \cite{leo} in his seminal 
analysis of the Maxwell's demon problem. The Brownian particle has a
Hamiltonian $H(p,x,t)$, where $p,x$ are coordinate and momentum. 
A parameter which vary with time can be the mass of the particle or
the shape of its potential energy.
The environment of the particle will be taken to be a thermal bath.
This is a generic situation, in the sense that the bath satisfies to
the following generally accepted conditions \cite{klim,balian,weiss}, 
which are exactly the same for both quantum and classical cases. 

1) The interaction between the particle and bath is linear. 
It is assumed to be so weak that the non-linear
modes of the bath are not excited, and it can be modeled itself 
as a collection of harmonic oscillators \cite{AN1,caldeira,weiss}.
This assumption has been verified rigorously, when starting from 
rather general microscopic situations.

2) The bath is a macroscopic system, namely the thermodynamical limit 
is assumed to be taken for it. 

3) Before to start to interact with the particle 
at some initial time the bath was in
equilibrium (i.e. in a Gibbsian state) at temperature $T$. 
This temperature will be refered to as the temperature of the bath. 
This assumption reflects the typical macroscopic preparation at the
initial time.

4) The particle and bath together form a closed system.
Thus, the overall system is described by the Schr\"odinger equation
(alternatively Heisenberg equations) in the quantum case, and by the Newton
equations in the classical case.

A minimal model, which incorporates all these properties was proposed
in \cite{ford,ullersma}, and much later became known as the 
Caldeira-Leggett model \cite{caldeira,weiss}.
The above assumptions ensure that the reduced dynamics of the Brownian 
particle will be given by the quantum or classical Langevin equations
\cite{weiss}.

As a result of interaction with the macroscopic bath, the Brownian 
particle will relax with  time towards a definite stationary state.
In the present paper we will additionally assume that all external 
operations on the particle
are adiabatic, namely they occur on time-scales which are much larger than 
the characteristic relaxation time. There are several physical reasons for
this restriction. First, in many circumstances the adiabatic process can be
shown to be optimal, in the sense that it is connected with minimal 
amount of work done by the external agent \cite{AN1,klim,balian}. On the 
other hand, this time-scale separation more
naturally corresponds to the interaction between a deterministic agent
and the microscopic particle.

Let us now consider the classical and quantum situation in separate.

\subsubsection{Classical case}

As well known under the above standard
assumptions on the thermal bath the classical 
Brownian particle relax to the Gibbs distribution
\cite{klim,balian,risken}:
\begin{eqnarray}
\label{w1}
&&P(p,x) = \frac{1}{Z }\exp [-\frac{1}{T}H(p,x)], \nonumber\\
&&Z=\int \d x~\d p ~\exp [-\frac{1}{T}H(p,x)]
\end{eqnarray}
Since the external operation is assumed to be adiabatic, the time-dependent 
distribution of the particle will be given by Eq.~(\ref{w1}) with the 
corresponding time-dependent Hamiltonian $H(x,p,t)$.
The Clausius equality
\begin{equation}
\label{C}
\dbarrm {\cal Q} = T\d S ,
\end{equation}
which connect the changes of heat and entropy during the process
can be derived directly from Eqs.~(\ref{w2}, \ref{w22}, \ref{dE}).
It holds that when compressing 
the phase space of the particle ($\d S<0$), it releases heat 
($\dbarrm{\cal Q}<0$). Since for any non-adiabatic change 
one has $\dbarrm {\cal Q} \leq T\d S$
(Clausius inequality), $|\dbarrm {\cal Q}|$ can only increase for not very 
slow processes. In other words, the minimal 
amount of the released heat is equal to $|T\d S|$. 
This is the Landauer principle.

\subsection{Quantum case}

Let us now move to the quantum domain, which in the present context
just means the domain of low temperatures. 
We {\it assume} that the quantum carrier of information interacts 
with its thermal bath, but so weakly, that it is described by 
quantum Gibbs distribution
at the bath temperature $T$:
\BEA
\label{q3}
\rho = \frac{1}{Z}\exp [-\frac{H}{T}], \qquad
Z={\rm tr }\exp [-\frac{H}{T}]. 
\EEA
The concrete conditions on the interaction strength will be discussed
later. Now it can be easily seen that provided we use entropy as defined in
Eq.~(\ref{q2}) the Clausius inequality (\ref{C}) is still holds and all its
consequences including the Landauer principle are generalized automatically.
The important difference between the classical and quantum cases has to
be noted already here: In the former situation there is no limitation on
the interaction strength, and the classical Gibbs distribution appears
naturally from the above standard conditions on the thermal bath.

\renewcommand{\thesection}{\arabic{section}}
\section{Quantum Brownian particle 
in contact with its thermal bath}
\setcounter{equation}{0}\setcounter{figure}{0} 
\renewcommand{\thesection}{\arabic{section}.}

\subsection{Wigner function and effective temperatures}

As explained in the introduction, at low temperatures of the bath 
the Brownian particle is not 
described by the quantum Gibbs distribution, except very weak 
interaction with the bath. Therefore, its state
at low temperatures has to be found from first principles starting from
microscopic description of the bath and the particle. This program
was realized in \cite{AN,AN1,weiss,haake}. In particular, in 
\cite{AN,AN1} we investigated statistical thermodynamics of the quantum 
Brownian particle.

Here we consider the simplest example of harmonic oscillator
with Hamiltonian 
\BEA
\label{ham}
H(p,x)=\frac{p^2}{2m}+\frac{ax^2}{2},
\EEA
where $m$ is the mass, and $a$ is the width.
The state of this particle can be described through the Wigner 
function \cite{balian}. Recall that in quantum theory this 
object plays nearly the same role as the common distribution 
of coordinate and momentum in the classical theory.
The stationary Wigner function reads \cite{grabert,weiss}\cite{AN,AN1}: 
\begin{eqnarray}
\label{ole77}
W(p,x)&&=W(p)~W(x)\nonumber \\
&&=\frac{1}{2\pi}\sqrt{\frac{a}{mT_pT_x}}\,\,
\exp{[-\frac{p^2}{2mT_p}-\frac{ax^2}{2T_x}]},
\end{eqnarray}
where $W(p)$, $W(x)$ are the probability distributions of
momenta and coordinate, and
\BEA
\label{T}
T_x=a\langle x^2\rangle , \qquad T_p=\langle p^2\rangle /m
\EEA
are effective temperatures to be discussed a bit later.
Eq.~(\ref{ole77}) represents the state
of the particle provided that the interaction with the bath was 
switched long time before, so that the particle already came to 
its stationary state.

The effective temperatures $T_p$ and 
$T_x$ depend not only on the system parameters $m$, $T$, and $a$,
but also on the damping constant $\gamma$, which quantifies
the interaction with the thermal bath, and on a large 
parameter $\Gamma$ which is the maximal characteristic
frequency of the bath. In particular, the Gibbsian limit
corresponds to $\gamma \to 0$. Then distribution
(\ref{ole77}) tends to the quantum Gibbsian:
$T_x=T_p=\half\hbar\omega_0{\rm coth}(\half\beta\hbar\omega _0)$,
where $\omega _0=\sqrt{a/m}$ (see Eqs.~(\ref{baran1}, \ref{baran1})). 
In the classical limit, which is realized for $\hbar\to 0$ or $T\to
\infty$, the dependence on $\gamma $ and $\Gamma $ disappears, and
$T_p$, $T_x$ go to $T$, reproducing the classical Gibbsian 
distribution (\ref{w1}). The appearance of the effective temperatures 
in the quantum regime can be understood as follows.
For $T\to 0$ quantum Gibbs distribution predicts the pure vacuum state
for the particle. Since due to quantum entanglement this is impossible
for the non-weakly interacting particle, $T_x$, $T_p$ depend on $\gamma $,
and have to be obtained from the first principles, since the state is not
any more Gibbsian.

The exact expressions for $T_p$, $T_x$ reads
\cite{grabert,weiss}\cite{AN1}
\BEA
T_p=\frac{\hbar \gamma\Gamma ^2}{\pi m}~
\frac{\omega _3^2\psi _3}
{(\omega _3^2-\omega _1^2)(\omega _3^2-\omega _2^2)}
+\frac{\hbar \gamma\Gamma ^2}{\pi m(\omega _1^2-\omega _2^2)}
\left [
\frac{\omega _2^2\psi _2}
{\omega _3^2-\omega _2^2}
-\frac{\omega _1^2\psi _1}
{\omega _3^2-\omega _1^2}\right ]-
T,
\label{wuk1}
\EEA
\BEA
T_x=-\frac{a\hbar \gamma\Gamma ^2}{m^2\pi}~
\frac{\psi _3}
{(\omega _3^2-\omega _1^2)(\omega _3^2-\omega _2^2)}
+\frac{a\hbar \gamma\Gamma ^2}{m^2\pi (\omega _2^2-\omega _1^2)}
\left [
\frac{\psi _2}
{\omega _3^2-\omega _2^2}
-\frac{\psi _1}
{\omega _3^2-\omega _1^2}\right ]-T,
\label{wuk2}
\EEA
where
$\psi _k=\psi[\hbar\beta\omega _k/(2\pi)]$, $k=1,2,3$, and 
$\psi (z)=\Gamma '(z)/\Gamma (z)$ is Euler's psi-function.
$\omega _{1,2,3}$ are roots of 
the following cubic equation
\BEA
(\Gamma -\omega  )(\omega ^2+\omega _0^2)-\omega
\frac{\gamma \Gamma }{m}=0,
\EEA
In the present paper we will be mostly interested by 
the so-called quasi-Ohmic limit where $\Gamma$ is the
largest characteristic frequency of the problem. 
This is the most realistic situation with realistic
information storing devices. In this limit one approximately has:
\BEA
&&\omega _{1,2}=\frac{\gamma}{2m}
\left (1\pm \sqrt{1-4\xi }\right )
+\frac{\gamma ^2 }{2\Gamma m^2}
\left[ 1\pm
\frac{1-2\xi}{\sqrt{1-4\xi}}
\right ],\nonumber\\
\label{mega1}
&&
\\
&& \omega _3=\Gamma -\frac{\gamma }{m}-\frac{1}{\Gamma}
\left (\frac{\gamma }{m}\right )^2
,
\label{mega2}
\EEA
where an important parameter $\xi = am/\gamma ^2$ characterizes
the relative importance of damping: $\xi \ll 1$ corresponds to 
the overdamped motion, whereas the converse case indicates underdamping.
We will basically use the first leading terms in 
Eqs.~(\ref{mega1}, \ref{mega2}). In this way we
obtain for the effective temperatures:
\begin{eqnarray}
\label{Tp}
&&T_p=\frac{\hbar}{\pi(\omega_1-\omega_2)}\left [
(\omega_1^2-\omega_2^2)\psi(\frac{\beta\hbar\Gamma}{2\pi})
-\omega_1^2\psi(\frac{\beta\hbar\omega_1}{2\pi})
+\omega_2^2\psi(\frac{\beta\hbar\omega_2}{2\pi})\right ]-T,\\
&&T_x=\frac{\hbar a}{m\pi(\omega _1-\omega _2)}
\left [\psi (\frac{\beta \hbar\omega _1}{2\pi})-
\psi (\frac{\beta \hbar\omega _2}{2\pi})\right ]-T,
\label{Tx}
\end{eqnarray}
Their derivation can be found in \cite{AN1}. 
Particular cases can be studied with help of the following
approximate values for $\psi $-function.
\begin{eqnarray}
\label{kumar1}
&&\psi (x)=-\frac{1}{x}-\gamma _{E}
+x\frac{\pi ^2}{6}, \quad |x|\ll 1
\\
\label{kumar2}
&&\psi (x)=\ln x-\frac{1}{2x}-\frac{1}{12x^2}, \quad |x|\ge 1
\end{eqnarray}
where $x$ is a complex number, and $\gamma _{E}=0.577216$ is 
the Euler constant.
In the low-temperature limit $T\to 0$ we obtain from 
Eqs.~(\ref{Tp}, \ref{Tx}):
\begin{eqnarray}
&&T_p=\frac{\hbar 
[\omega_1^2\ln\frac{\Gamma}{\omega_1}
-\omega_2^2\ln\frac{\Gamma}{\omega_2}]
}{\pi (\omega_1-\omega_2)}+{\cal O}(T^4),
\label{Tp1}
\end{eqnarray}
\BEA 
&&T_x=
\frac{\hbar a}{\pi m(\omega_1-\omega_2)}
\ln\frac{\omega_1}{\omega_2}+\frac{\pi\gamma}{3\hbar a}T^2 +{\cal O}(T^4)
\label{Tx1}
\end{eqnarray}
The weak-coupling limit can be obtained from
Eqs.~(\ref{Tx}, \ref{Tp}) taking $\xi \gg 1$ and noticing
\BEA
\frac{\psi (ix)-\psi (-ix)}{i\pi}=\frac{1}{x\pi} +{\rm coth}(x\pi ),
\EEA
which is obtained with the reflection formula:
$\Gamma (z)\Gamma (1-z)=\pi/(\sin \pi z)$. Here one has the following 
expressions:
\BEA
\label{baran1}
&&T_p=\frac{\hbar\omega _0}{2}{\rm coth}\frac{\hbar\beta\omega _0}{2}
+ \frac{\hbar\gamma}{4\pi m}{\cal G}(\frac{i\hbar\beta\omega _0}{2})
+\frac{\hbar\gamma}{4\pi m}\left [
2\psi(\frac{\beta\hbar\Gamma}{2\pi})-\psi(\frac{i\beta\hbar\omega _0}{2\pi})-
\psi(-\frac{i\beta\hbar\omega _0}{2\pi})
\right ],\\
\label{baran2}
&&T_x=\frac{\hbar\omega _0}{2}{\rm coth}\frac{\hbar\beta\omega _0}{2}
+ \frac{\hbar\gamma}{4\pi m}{\cal G}(\frac{i\hbar\beta\omega _0}{2}),\\
&&{\cal G}(x)=\frac{x}{i}[\psi '(ix)-\psi '(-ix)]
\EEA
The asymptotic expressions 
of these quantities  read in the opposite, strongly damped region 
$\xi \ll 1$ and for low $T$:
\begin{eqnarray}
\label{Tp0}
&&T_p=\frac{\hbar \gamma }{\pi m}\ln\frac{\Gamma m}{\gamma}+
\frac{\hbar a}{\pi \gamma}+{\cal O}(T^4), \\ 
\label{Tx0}
&&T_x=\frac{\hbar a}{\pi\gamma}\ln\frac{\gamma^2}{am}
+\frac{\pi\gamma}{3\hbar a}T^2 +{\cal O}(T^4),
\end{eqnarray}
It is interesting to mention as well the high-temperature (quasi-classical)
asymptotic values for $T_p$, $T_x$. Applying Eq.~(\ref{kumar1}) one gets
\begin{eqnarray}
\label{highTp}
&&T_p= T+\frac{\hbar ^2(am-\gamma ^2+\Gamma m\gamma )}{12m^2T}+{\cal O}(\hbar ^3\beta ^2), \\
&&T_x= T+\frac{\hbar ^2a}{12mT}+{\cal O}(\hbar ^3\beta ^2)
\label{highTx}
\end{eqnarray}

\subsection{Energy and partial entropies}
The average energy of the Brownian particle
\BEA
\label{w7}
U=\int \d x\d p W(p,x)H(p,x)=\frac{T_p}{2}+\frac{T_x}{2}
\EEA
does depend on $a$ and $m$, in contrast to its classical value $T$.
We will need entropies of momentum and coordinate distribution
\begin{eqnarray}
\label{S}
&&S_p= -\int \d p W(p)\ln W(p)=\half\ln (mT_p), \\
&&S_x=-\int \d x W(x)\ln W(x)=\half\ln \frac{T_x}{a}
\label{S1}
\end{eqnarray}
The `Boltzmann' entropy reads
\BEA S_B=-\int\d p\d x\,W(p,x)\ln[\hbar W(p,x)]
=S_p+S_x-\ln\hbar=\half\ln\frac{mT_pT_x}{a\hbar^2}\EEA
Notice that they all
are different from $\SN (\rho )$ defined by (\ref{q2}).

\subsection{Heat and work}

The expressions for heat and work are generalized from Eqs.~(\ref{dE})
by simply using the Wigner function $W(p,x)$ instead of $\rho$.
This can be easily verified, when using Eq.~(\ref{korkud}).
One can prove by a direct calculation that quantities $T_p$, $T_x$
do deserve their nomenclature, since the classical Clausius equality can 
be generalized as
\begin{eqnarray}\label{asala2}
\d U=\dbarrm {\cal Q}+\dbarrm {\cal W}=
T_p\d S_p+T_x\d S_x+\dbarrm {\cal W}
\end{eqnarray}
for variation of any parameter. 

We will be especially interested in variation of the mass and the width
of the potential. The corresponding changes of heat read
\BEA
\label{khosrov}
\dbarrm_a{\cal Q}=
\half\left(\frac{\p T_p}{\p a}+\frac{\p T_x}{\p a}-\frac{T_x}{a}\right)\d a,
\EEA
\BEA
\label{khosrov00}
\dbarrm_m{\cal Q}=
\half\left(\frac{\p T_p}{\p m} + \frac{\p T_x}{\p m}+\frac{T_p}{m}\right)\d m,
\EEA
Using Eqs.~(\ref{Tx}, \ref{Tp}) one can get general formulas 
for heat. Let us first introduce following notations
\BEA
\label{khorom1}
z=\sqrt{1-\xi },\qquad \alpha _1 =\frac{\hbar \gamma}
{4\pi mT}, \qquad
\alpha _2 =\frac{a \hbar }{\pi\gamma T},
\EEA
and get
\begin{eqnarray}
\label{Qa}
\frac{\p{\cal Q}}{\p a}=
\frac{2mT}{\gamma ^2}
\left (\frac{1}{1-z^2} 
+\frac{\alpha _1^2}{z}
\{ (1+z)\psi ' (\alpha _1 [ 1+z]  ) - (1-z)\psi ' (\alpha _1 [ 1-z]) 
\}
\right ),
\end{eqnarray}
\begin{eqnarray}
\label{Qm}
\frac{\p{\cal Q}}{\p m}=
\frac{2aT}{\gamma ^2}
\left (\frac{1}{z^2-1}
+\frac{\alpha _2^2}{z}
\{ \frac{1}{(1-z)^3}\psi ' (\frac{\alpha _2}{  1-z}  ) - 
\frac{1}{(1+z)^3}\psi ' (\frac{\alpha _2}{  1+z}  ) \}\right )
\end{eqnarray}
Following to asymptotic expressions of $\psi '(x)$ given by 
Eqs.~(\ref{kumar1}, \ref{kumar2}) one derives
\begin{eqnarray}
\label{comrad0}
&&\frac{\p{\cal Q}}{\p a}
=-\frac{T}{2a}+\frac{\hbar ^2}{24mT}, \qquad \alpha _1\ll 1
\\
&&\frac{\p{\cal Q}}{\p a}
=-\frac{\pi\gamma  T^2}{3\hbar a^2}, \qquad \alpha _1\ge 1 
 \label{khorosh} \\
&&\frac{\p{\cal Q}}{\p m}
=\frac{T}{2m}+\frac{\hbar ^2\gamma ^2(z^2+1)}{66m^3T}, \qquad \alpha _2\ll 1
\\
&&\frac{\p{\cal Q}}{\p m}
=\frac{\hbar\gamma  }{2\pi m^2}, \qquad \alpha _2\ge 1 
\label{comrad}
\end{eqnarray}
Notice that the last equation applies not only for low temperatures, but
also for weak coupling (see Eq.~(\ref{khorom1})).

Using these results one can show that
\BEA
\label{bogemia}
\frac{\p{\cal Q}}{\p a}\le 0,
\qquad
\frac{\p{\cal Q}}{\p m}\ge 0
\EEA
for all values of parameters including, of course, the classical limit.
For the work done in this processes one obtains
\BEA
\label{khosrov2}
\frac{\p {\cal W}}{\p a}=\half\langle x^2\rangle
=\half \frac{T_x}{a}\ge 0,
\EEA
\BEA
\label{khosrov002}
\frac{\p {\cal W}}{\p m}
=-\half\frac{\langle p^2\rangle}{m^2}
=-\half \frac{T_p}{m}\le 0
\EEA
It is interesting to mention that the signs of $\partial {\cal Q}$ and
$\partial {\cal W}$ in Eqs.~(\ref{bogemia},
\ref{khosrov2}, \ref{khosrov002}) are the same as in the classical case,
where $T_x=T_p=T$.

\subsection{Density matrix}

To investigate von Neumann entropy (\ref{q2}) one needs density
matrix corresponding to the Wigner function (\ref{ole77}). 
Applying the standard relation:
\BEA
\langle x+\frac{u}{2}|\rho | x-\frac{u}{2} \rangle =
\int \d p~e^{-ipu/\hbar}W(p,x)
\label{korkud}
\EEA
which connects the density matrix in coordinate representation
with the Wigner function, one gets the following expression
\begin{eqnarray}
\label{den2}
\langle x|\rho | x' \rangle &&=
\frac{1}{\sqrt{2\pi \langle x^2\rangle }}
\exp{\left [-\frac{(x+x')^2}{8\langle x^2\rangle}
-\frac{(x-x')^2}{2\hbar ^2/\langle p^2\rangle} \right ]}
\end{eqnarray}
The physical meaning of Eq.~(\ref{den2}) is clear: The diagonal
elements ($x=x'$) are distributed at the scale
$\sqrt{\langle x^2\rangle}$,
while the maximally off-diagonal elements
($x=-x'$), which characterize
coherence, are distributed with the characteristic scale
$\hbar/\sqrt{\langle p^2\rangle}$.

We have to find eigenfunctions and eigenvectors
of this density matrix
\BEA
\int \d x' \langle x|\rho | x' \rangle f_n(x')
=p_n f_n(x)
\EEA
The solution of this problem uses some tabulated 
formulas for Hermite polynoms, and results in
\begin{eqnarray}
&&p_n = \frac{1}{w+\half}\left [ 
\frac{w-\half}{w+\half}
\right ]^n, \\
&&f_n(x) = c\,H_n(c\,x)e^{-c^2x^2/2}, \qquad c=
\left(\frac{\langle p^2\rangle}{\hbar ^2\langle x^2\rangle}
\right)^{1/4} \\
&&w=\frac{\Delta p\,\Delta x}{\hbar}=
\sqrt{\frac{\langle p^2\rangle\langle x^2\rangle}{\hbar ^2}}=
\sqrt{\frac{mT_pT_x}{\hbar ^2a}} \label{w=}
\end{eqnarray}
where $H_n$ are Hermite polynomials, and it holds that $w\ge \half$
due to the Heisenberg uncertainty relation. 
The result for the von Neumann entropy (\ref{q2})
now reads~\cite{weiss}
\BEA\label{SvN=}
S_{vN} =(w+\half)\ln(w+\half) -(w-\half)\ln(w-\half)
\EEA
The first terms in its large $w$-expansion are 
\BEA
\label{katmandu}&&
S_{vN} 
=\ln w+1-\frac{1}{24w^2}-\frac{1}{320w^4}-\frac{1}{2688w^6}
\EEA

Notice that the same quantity $w$ governs the Boltzmann entropy 
\BEA
S_B=S_p+S_x-\ln\hbar
=\ln{w}+1
\EEA
This appears to coincide with the leading terms of (\ref{katmandu}).
It is known to be larger than the Von Neumann entropy, and this is 
obvious from the sign of the correction terms.

If some parameter ($a$ or $m$) is varied, then the derivative
of $S_{vN}$ with respect to it reads:
\begin{eqnarray}
\d S_{vN}= \ln\frac{w+\half}{w-\half}\,\d w
\label{mustafa}
\end{eqnarray}
In other words, the sign of the change in $S_{vN}$ is determined by the sign
of the change in $w$. This holds as well for the change in $S_B$, so
qualitatively they carry the same information.

In this context
let us stress again that von Neumann entropy $\SN (\rho )$ is the 
unique quantum measure of localization and information, whereas the 
entropies $S_p$, $S_x$ characterize localizations of momenta and coordinate 
separately. Differences between $S_p+S_x$ and $\SN$ are due to the
fact that in quantum theory momentum and coordinate cannot be measured 
simultaneously; in this sense $S_p+S_x$ characterize two different 
measurement setups. Nevertheless, for the harmonic particle if
$\SN$ increases (decreases), then $S_p+S_x$ increases (decreases)
as well. Notice that the real importance of $S_p$, $S_x$ becomes
clear when they have to be used to generalize the Clausius inequality.
The von Neumann entropy $\SN$ cannot be used for this purpose
if $T_x\not =T_p$.

\renewcommand{\thesection}{\arabic{section}}
\section{Entropy decrease with heat adsorption}
\setcounter{equation}{0}\setcounter{figure}{0} 
\renewcommand{\thesection}{\arabic{section}.}

Now we will show that there are erasure processes, namely processes where
$\d \SN\le 0$, which are accompanied by adsorption of heat. As
we know heat is always adsorbed, when the mass is increased 
(see Eq.~(\ref{comrad})). It will be shown 
that there is a mass-increasing process, where $\d \SN\le 0$.
Using Eqs.~(\ref{Tx0}, \ref{Tp0}) one has
for $\partial _m(\langle x^2\rangle \langle p^2\rangle )$
at very low temperatures:
\begin{eqnarray}
\label{werner}
&&\frac{\partial w^2 }{\partial m}=\frac{\partial }{\partial m}\left [
\frac{1}{\hbar ^2}
\langle x^2\rangle \langle p^2\rangle \right ]\nonumber\\
&&= \frac{a}{\pi ^2\gamma ^2}
\left [
-1-\frac{\gamma ^2}{am}\ln \frac{\Gamma m}{\gamma}
+(1+ \frac{\gamma ^2}{am})\ln \frac{\gamma ^2}{am}
\right ]
\end{eqnarray}
This expression is negative in its range of applicability.

\begin{figure}[bhb]
\vspace{0.1cm}\hspace{-1.5cm}
\vbox{\hfil\epsfig{figure= 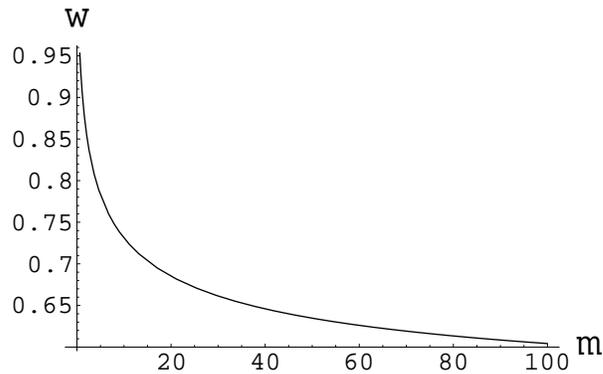,width=8cm,angle=0}\hfil}  
\vspace{0.75cm}
\caption{ Dimensionless phase space volume 
$w=\Delta p\Delta x/\hbar=
\sqrt{\langle x^2 \rangle \langle p^2 \rangle/\hbar ^2}$, 
versus mass $m$. The other parameters are
$a=\gamma =1$, $\Gamma =500$, $\hbar =1$ and
$T=0$. It is seen that the volume decays 
monotonically towards its minimal value $1/2$, 
set by the uncertainty relation.}
\label{fig1}
\end{figure}

An analogous argument can be brought about in the weak-coupling 
case. Having started from Eqs.~(\ref{Tp1}, \ref{Tx1}) or alternatively
from Eqs.~(\ref{baran1}, \ref{baran2}) one derived the following expressions
for the effective temperatures in the weak-coupling $\gamma\to 0$ and
low-temperature limit:
\BEA
\label{ahmad1}
&&T_p=\frac{\hbar\omega _0}{2}+\frac{\hbar\gamma}{\pi m} 
\ln\frac{\Gamma }{\omega _0\sqrt{e}},\\
\label{ahmad2}
&&T_x=\frac{\hbar\omega _0}{2}-\frac{\hbar\gamma}{2\pi m} 
\EEA
This implies
\BEA
\frac{\partial w^2 }{\partial m}
=-\frac{\gamma}{4m\sqrt{am}}\ln\frac{\Gamma }{\omega _0 e^2},
\EEA
which is again negative in its range of applicability
$\Gamma \gg \omega _0$.
The general situation at low temperatures is illustrated by
Fig.~\ref{fig1}, where it is seen that 
at low temperatures the dimensionless phase-space 
volume $w=\sqrt{\langle x^2 \rangle \langle p^2 \rangle/\hbar ^2}$ 
monotonically decreases when increasing the mass. 
In the limit $m\to\infty$ it tends to its corresponding gibbsian
value. This can be understood noticing that 
the stationary state of a very heavy Brownian
particle will not be influenced much by the bath.
Indeed, as seen from Eqs.~(\ref{mega1}, \ref{mega2}, \ref{Tp}, \ref{Tx})
the dimensionless parameter which controls transition from the
weakly damped to the strongly-damped regime is $\xi =am/\gamma ^2$.
So to increase the mass for all other parameters 
are being fixed produces the same effect
as decrease of the coupling constant $\gamma$.

Recall that the corresponding expression
(\ref{Qm}) for $\p {\cal Q}/\p m$ was 
positive. This just means that for the variation of $m$ we have an 
interesting case where heat is adsorbed when entropy is decreasing.
This is a counterexample
for the general validity of the Landauer principle.

\subsection{Where classical intuition is correct and
where it fails}

In the 
classical case one has an intuitively clear result: Upon increasing
the mass of the particle its entropy increases and it releases heat.
At the same time it does work against the external agent 
($\p {\cal W}/\p m <0$). 

The first part of this result holds as well in the quantum case, 
as follows from Eq.~(\ref{khosrov002}). However in the first part 
a unusual point appears: The quantum particle decreases its 
entropy when the mass is increased. Simultaneously, it adsorbs heat.
To understand this point we notice that at low temperatures of the bath
the particle has an appreciable entropy due to entanglement. When its mass
is increased, its state moves towards the gibbsian limit, and the entropy is
reduced just because in the zero-temperature gibbsian case the entropy is
just zero. This will take also for low but finite 
temperatures of the bath, as far as entanglement 
appreciable contribute to the entropy. So it is entanglement which leads 
to such a counterintuitive result.

Notice that this effect does not imply a violation of the second
law in Thomson's formulation, which speaks about 
the impossibility to extract work by a cyclic variation 
of a system parameter \cite{AN,AN1}.
Indeed, if, after increasing the mass,  one decrease it
in order to complete the cycle,
the external agent will do work on the particle, 
and it will release heat, thereby
nullifying the overall work and heat (as expected,
the overall work is positive if
non-adiabatic variations are considered \cite{AN,AN1}).

\subsection{Finite temperatures}

\begin{figure}[bhb]
\vspace{0.1cm}\hspace{-1.5cm}
\vbox{\hfil\epsfig{figure= 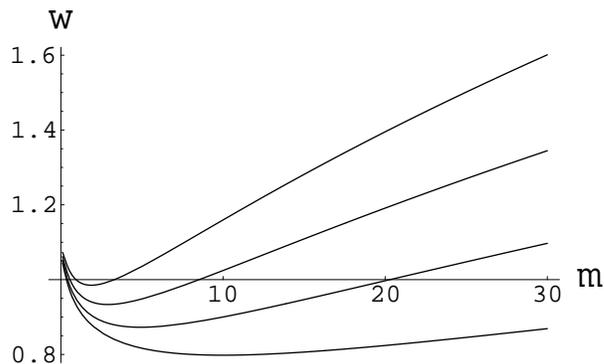,width=8cm,angle=0}\hfil}  
\vspace{0.75cm}
\caption{ Dimensionless phase space volume 
$w=\Delta p\Delta x/\hbar=
\sqrt{\langle x^2 \rangle \langle p^2 \rangle/\hbar ^2}$, 
versus mass $m$. From the top to the bottom: $T=0.25,0.20,0.15,0.10$.
The other parameters are the same as in Fig.~\ref{fig1}:
$a=\gamma =1$, $\Gamma =500$, $\hbar =1$. 
It is seen that there is a region of $m$'s where the volume decays. 
This region is completely shrunk for $T=0.47553$. For higher temperatures
the phase space volume monotonically increases with $m$.}
\label{fig2}
\end{figure}

The above effect $\partial w/\partial m <0$ was 
analytically illustrated for $T\to 0$.
However, it persists as well at finite, but sufficiently small temperatures.
This situation is illustrated in Fig.~(\ref{fig2}). Since in the classical
case, namely with high temperatures, one always has $\partial _m w
=T/(2\sqrt{ma})>0$, we expect that the region with $\partial _m w<0$ will
completely disappear at some finite temperature. This is indeed the case
as Fig.~(\ref{fig2}) shows.

\subsection{Variation of the spring constant}

The analogous variation of $a$ does not lead to such an usual result.
Here instead of Eq.~(\ref{werner}) one has
\begin{eqnarray}
\label{werner1}
&&\frac{\partial }{\partial a}\left [\frac{1}{\hbar ^2}
\langle x^2\rangle \langle p^2\rangle \right ]\nonumber\\
&&= \frac{m }{\pi\hbar \gamma ^2}
\left [
-1-\frac{\gamma ^2}{am}\ln \frac{\Gamma m}{\gamma}
+\ln \frac{\gamma ^2}{am}
\right ]\le 0,
\end{eqnarray}
but the corresponding expression (\ref{Qa}) for $\p {\cal Q}/\p a$
is negative as well.
An expression analogous to Eq.~(\ref{werner}) can be gotten also in 
the weak-coupling limit. Using Eqs.~(\ref{ahmad1}, \ref{ahmad2}) one 
gets:
\BEA
\frac{\partial w^2 }{\partial a}
=-\frac{\gamma}{a\sqrt{am}}\ln\frac{\Gamma }{\omega _0 }<0
\EEA
An important fact should be mentioned here. Although
the particle just releases heat during localization, 
this heat scales at low temperatures as $|\dbarrm{\cal Q}|\sim T^2\d a$ 
(see Eq.~(\ref{khorosh})). This is already
invalidating the Landauer bound 
$|\dbarrm{\cal Q}|\ge T|\d S|\sim T\d a$.

In this context the parameters of a statistical system can be 
distinguished as active and passive. In our concrete case the width
of the potential $a$ is a passive parameter, in a sense that its 
variations result in effects, which for any temperature are
qualitatively (but not quantitatively) similar to the
classical case. In contrast, the active parameters (in our
case it is the mass $m$) invert their behavior at low temperatures.
If one needs to increase entropy with adsorption of heat he/she
is advised to vary a passive parameter. In contrast, being aimed to
adsorb heat during contraction of entropy, one varies an active parameter.

\subsection{Weak coupling limit}
Here we will especially point out on applicability of our result in
the weak-coupling limit. First we will make an obvious remark
that the precise meaning of
this limit {\it must not} be understood in the sense $\gamma\equiv 0$,
since the damping constant $\gamma$ is never explicitly zero in practice, 
and having put it zero one will not have at all a possibility to change 
entropy of the particle. The weak -coupling limit is understood in a sense
that the interaction energy of the particle and the bath happened to be
sufficiently small compared to the energy of the particle itself \cite{balian}
(since the energy of the bath is infinite there is no need to involve
it here). For low temperatures and $\gamma\to 0$ the energy of the particle 
is given by its zero-point value which is $\half\hbar\sqrt{a/m}$. Because
of the interaction energy is explicitly zero for $\gamma\equiv 0$, it will
be enough to choose $\gamma$ sufficiently small to ensure the above condition
of the weak-coupling limit. 

Let us now turn to Eq.~(\ref{comrad}) which represents the 
amount of heat obtained by the particle when changing the mass at low 
temperatures. It is seen that in the leading order this quantity is 
proportional to $\gamma$. Thus, although the particle is 
in the weak-coupling regime, it still gets a positive 
(though small) amount of heat during variation of its mass.

\subsection{High temperatures}

Finally we wish in more details that the Landauer principle does hold in our 
model for sufficiently high temperatures. This follows fron the fact that in
this limit $T_p,T_x\to T$ as seen from Eqs.~(\ref{highTp}, \ref{highTx}).
A more elaborated discussion goes as follows. One has the following exact 
relation:
\begin{eqnarray}
\label{karavan}
\dbarrm {\cal Q}-T\d \SN =(T_x-T)\d S_x
 +(T_p-T)\d S_p
-T\d (\SN -S_x-S_p)
\end{eqnarray}
One applies here Eqs.~(\ref{highTp}, \ref{highTx}, \ref{katmandu})
to get for variation of $m$
\begin{eqnarray}
\label{murad}
\frac{\p {\cal Q}}{\p m}-T\frac{\p \SN}{\p m}
=-\frac{\hbar ^2\gamma ^2}{24m^3T}+{\cal O}(\hbar ^3\beta ^2)
\end{eqnarray}
Now it is seen that the deviation from the Clausius equality,
and thus from the Landauer principle, will disappear for high 
temperatures or for $\gamma\to 0$ and/or $\hbar\to 0$ as should be. 
It is seen as well that the correction
has the opposite sign to the main effect: When increasing mass
the particle releases heat as the main standard thermodynamical
effect, but the small
correction in r.h.s. of Eq.~(\ref{murad}), which appears due to the common
influence of the quantum effects and interaction with the bath, tends
to do this released heat slightly smaller.

\renewcommand{\thesection}{\arabic{section}}
\section{On a popular derivation of the Landauer bound}
\setcounter{equation}{0}\setcounter{figure}{0} 
\renewcommand{\thesection}{\arabic{section}.}

Let us discuss in a more general perspective the obtained result on 
the violation of the Landauer principle.
For this purpose we will analyze one of the simplest derivations
of this principle \cite{landauer,landauer1,ben}, in order 
to understand what essentially goes into it
and where its argument may be inapplicable. 
The derivation goes as follows.
Erasure is accompanied by reduction of entropy of the 
information-carrying system. Since entropy of the overall system,
which is the carrier plus bath, cannot decrease, one quickly concludes  
that entropy of the bath should increase thereby producing heat.
This argument seems to be rather solid, because, instead involving
any derivation, it just directly refers to the second law.
However, there are  {\it three assumptions} in it, 
which are rather restrictive and need not be
valid in situations of physical interest. 
The first assumption is that
the total entropy $S$ of the overall system is sum of 
partial entropies of system and bath, $S=S_S+S_B$. 
The second is quick thermalization
in the bath, implying $\dbarrm \Q_B=T\d S_B$. The third assumption
is smallness of the interaction energy $\dbarrm \Q_I$, allowing to 
conclude from the energy conservation $\dbarrm Q_S +\dbarrm Q_B+
\dbarrm Q_I=0$ that $\d S_B=\dbarrm \Q_B/T=-\dbarrm \Q_S/T$.
With these assumptions it now follows immediately that
$0\le \d S=\d S_S+\d S_B=\d S_S-\dbarrm \Q_S/T$.
These asumptions are strictly valid only for non-interacting information
carrier and its bath. However, without interaction there is no reason to
speak about erasure. These assumptions may be valid as certain
{\it approximations} in the weak coupling case plus several additional 
conditions \cite{balian}. Their validity is especially endangered
in the quantum regime where the complete entropy,
which is the subject of the second law applied to the complete system, 
is not equal to the sum of the separate entropies if there occurs 
quantum entanglement. 
So the above simple derivation is actually rather
restricted, as was noted already in the context of rather
different physical arguments \cite{berger,wolpert}.
It need not be invoked at all, 
since the Landauer principle can be completely embedded in the 
Clausius inequality. The latter is typically  valid 
without any weak-coupling limit, as it happens for classical
Brownian motion. Even in situations very far from the equilibrium 
thermodynamics, one can figure out regions of its validity \cite{N1,AN0}.
The conclusion of this brief consideration is that in situations
where the weak-coupling assumption is valid, 
the argument that the total entropy cannot decrease {\it need not be}
invoked, since it amounts to rederiving the Clausius inequality.
It can, of course, still be kept as a useful explanation.
But the general validity of the Landauer principle must be 
completely put on the the Clausius inequality; it is just a
direct consequence of this inequality, which has a substantially larger
validity than weak coupling.

Nevertheless, it was discussed here that quantum entanglement 
limits the validity of the Clausius inequality 
and, consequently, Landauer's bound.
It can be checked explicitly that 
in that regime all three above assumptions are invalid
~\cite{AN,AN1}. Recenly violations of other formulations of the second law
were noticed and investigated in \cite{c0,c1}.

\renewcommand{\thesection}{\arabic{section}}
\section{Conclusion}
\setcounter{equation}{0}\setcounter{figure}{0} 
\renewcommand{\thesection}{\arabic{section}.}

The Landauer principle requires dissipation (release) of $T|\d S|$ units
of energy as a consequence of erasure of $|\d S|$ units of information.
This was believed to be the only {\it fundamental} energy cost of
computational processes \cite{landauer,landauer1,ben,japan}. 
Though in practice computers dissipate much more
energy, the Landauer principle was considered to put a general physical
bound to which every computational device interacting with its thermal 
environment must satisfy. Indeed, in several physical situations the Landauer
principle can be proved explicitly \cite{japan}. 

The main purpose of the present paper was to provide a counterexample of 
this principle, and thus to question its univeral validity. In the reported 
case all general requirements on the information carrier and its interaction 
with the bath are met. The only new point of our approach is that we were
interested by sufficiently low temperatures, where quantum effects 
are relevant.
The Landauer principle appeared to be violated by these effects (in particular,
by entanglement). At high temperatures we reproduce its validity. In fact,
in this limit our model is equivalent to that considered in Ref.~\cite{japan},
where the classical Landauer principle was derived in a quite general ground.

Recently the Landauer bound attracted a serious attention by workers
in the field of applied information science \cite{ralf}. 
There is a definite belief that this bound
can be approached by further miniaturization of computational devices.
It is hoped that the present paper will help to understand limitations of
the Landauer principle itself, which may lead to unexpected
mechanisms for computing in the quantum regime.

\acknowledgments

A.E. A was supported by NATO and by FOM (The Netherlands).


\end{document}